%% file: ICC_Paper_V7.tex
\documentclass[conference,a4paper,10pt]{IEEEtran}
\usepackage{xcolor}
\usepackage{balance}
\usepackage{subcaption}
\usepackage{amsmath}
\usepackage{amsmath,amssymb,amstext}
\usepackage{bm}
\usepackage{algorithm}
\usepackage{algpseudocode}
\usepackage{comment}
\usepackage{caption} 
\usepackage{cite}

\ifCLASSINFOpdf
  \usepackage[pdftex]{graphicx}
\else
  \usepackage[dvips]{graphicx}
\fi

\ifCLASSOPTIONcompsoc
 \usepackage[caption=false,font=normalsize,labelfont=sf,textfont=sf]{subfig}
\else
 \usepackage[caption=false,font=footnotesize]{subfig}
\fi

\usepackage{pgfplots}
\pgfplotsset{compat=newest}
\usepgfplotslibrary{groupplots} 
\usepackage{tikz}
\usetikzlibrary{calc}
\usetikzlibrary{positioning}
\usetikzlibrary{arrows.meta}
\usetikzlibrary{spy}

\newlength{\figureheight}
\newlength{\figurewidth}

\colorlet{plot1}{cyan!30!white}
\colorlet{plot2}{red!40!white}
\colorlet{plot3}{cyan!70!black}
\colorlet{plot4}{black}

\newcommand{\ist}{\hspace*{.3mm}}
\newcommand{\rmv}{\hspace*{-.3mm}}
\newcommand{\iist}{\hspace*{1mm}}

\newcommand{\nn}{\nonumber}
\newcommand{\T}{\mathrm{T}}
\renewcommand{\H}{\mathrm{H}}

\begin{document}
%
\title{


Bayesian Self-Calibration and Parametric Channel Estimation for 6G Antenna Arrays


}

\author{\IEEEauthorblockN{
Patrick Hödl,
Jakob M\"oderl,
Erik Leitinger and
Klaus Witrisal
}                                     
\IEEEauthorblockA{
Institute of Communication Networks and Satellite Communications, TU Graz, Austria}
\IEEEauthorblockA{
\{patrick.hoedl, jakob.moederl, erik.leitinger, witrisal\}@tugraz.at}
}

\maketitle

\begin{abstract}
Accurate channel estimation is essential for both high-rate communication and high-precision sensing in 6G wireless systems. However, a major performance limitation arises from calibration mismatches when operating phased-array antennas under real-world conditions. To address this issue, we propose to integrate antenna element self-calibration into a variational sparse Bayesian learning (VSBL) algorithm for parametric channel estimation. We model antenna gain and phase deviations as latent variables and derive explicit update equations to jointly infer these calibration parameters and the channel parameters; the number of multipath components (MPCs) along with their complex amplitudes, delays, and angles-of-arrival (AoA), as well as the noise variance.
We assess its performance in terms of the optimal subpattern-assignment (OSPA) metric, demonstrating consistent improvements over conventional VSBL without calibration. 
Furthermore, we show that integrating the estimation of the calibration parameters into the VSBL algorithm actually increases convergence speed, since a missing or wrong calibration results in the additional estimation of spurious components.
\end{abstract}

\vskip0.5\baselineskip

\section{Introduction}
Channel estimation refers to the extraction of parameters that characterize a radio channel, from channel sounding measurements such as channel impulse responses (CIRs). From a signal processing perspective, this task can be formulated as a line spectral estimation (LSE) problem~\cite{BadHanFle:TSP2017,HanFleuRao:TSP2018,GreLeiWitFle:TWC2024}, where one aims to estimate the parameters of dictionary atoms parameterized by complex exponentials. The estimated parameters describe the multipath components (MPCs) present in the radio channel. These MPCs are caused by reflections in the environment and are expected to be highly resolvable in wideband 6G systems. To cope with this challenge and maximize the link budget at the receiver, the use of phased-array antennas becomes essential. However, phased arrays often exhibit unknown element gain and phase errors that can vary over time or with changing ambient conditions \cite{WuSanKesGusAmaWymTWC2025, TiaZhaWymTVT2025}. Such imperfections significantly degrade the accuracy and computational efficiency of channel estimation algorithms. Calibration methods, such as effective aperture distribution function (EADF) calibration, are typically performed using labor-intensive anechoic-chamber measurements~\cite{LanGal:EECWT2004,CaiXueZhuMeiFedAleTuf:TAP2023}. This process is time-consuming and impractical for user devices operating in dynamic scenarios. 

\subsection{State of the Art}

To overcome these limitations, gain and phase errors of individual antenna elements can be estimated jointly with the channel parameters. Approaches that integrate antenna element gain and phase error estimation directly into the channel estimation framework are still rare. Previous works have explored variants of sparse Bayesian learning (SBL) for robust wideband channel estimation with super-resolution capability. However, most approaches assume perfectly calibrated antenna arrays and treat calibration separately without integrating it into the signal model~\cite{TheRonKou:TSP2014,TonCheDenHu:Elec2025}. In~\cite{LiZha:PiEMR2022}, array gain and phase self-calibration is performed after obtaining coarse angle-of-arrival (AoA) estimates from an SBL algorithm using a narrowband signal model factorized by singular value decomposition. The calibration parameters are estimated by optimizing a cost function based on signal and noise subspace orthogonality, and the process is repeated until convergence. In~\cite{CheMaCheCao:IEEE2020}, phase error estimation is integrated into an SBL-based AoA estimation framework, where the phase error is modeled as a latent variable with a hierarchical prior. Again, a narrowband signal model is employed. A VSBL method with self-calibration is presented in~\cite{WanGuiZhaFuLiu:RS2025}, adopting a non-uniform noise narrowband signal model. Gain and phase errors are modeled as hyperparameters and point-estimated within an expectation-maximization (EM) scheme. While these works demonstrate the potential of SBL-based frameworks for joint AoA estimation and array calibration, they share a critical limitation: none of them employs a wideband signal model, which is indispensable for mmWave propagation. Furthermore, they rely on EM-based SBL approaches, which often leads to slow convergence and limits their practicality \cite{GreLeiWitFle:TWC2024}. Addressing these issues is essential to meeting the stringent accuracy and reliability requirements of emerging 6G wireless communication and sensing systems.

\subsection{Contributions}

Unlike purely data-driven approaches~\cite{RamHagFurMagWym:TWC2025_HMLISAC}, this paper proposes a model-based Bayesian framework for joint wideband channel estimation and array self-calibration, in which antenna gain and phase mismatches are explicitly integrated into the signal model and inferred jointly with the channel parameters. The main contributions are as follows.
\begin{itemize}
  \item We propose a hierarchical Bayesian model that embeds antenna element gain and phase errors in a wideband signal model as latent complex-valued calibration weights.

  \item We derive closed-form variational updates for the calibration parameters using a fast VSBL rule~\cite{ShuBucKulPoo:TSP2011}, enabling efficient joint inference with accelerated convergence.

  \item We demonstrate through numerical evaluations consistent performance gains over methods without joint calibration.
\end{itemize}
In contrast to \cite{WanGuiZhaFuLiu:RS2025}, which treats calibration parameters as deterministic unknowns, we model them as latent random variables considering uncertainty. Furthermore, instead of the classical EM-based SBL update used in \cite{WanGuiZhaFuLiu:RS2025}, we employ a fast SBL update rule~\cite{ShuBucKulPoo:TSP2011}, yielding improved convergence.


\section{Signal Model}

We consider a static wideband mmWave single-input multiple-output (SIMO) setup consisting of a single omnidirectional transmit antenna and a receiver equipped with a planar array comprising $P$ elements.
\footnote{The proposed model can be straightforwardly generalized to arbitrary three-dimensional array geometries and extended to MIMO configurations.}

\subsection{Continuous-Time Model}

The propagation environment is modeled by $K$ specular multipath components (MPCs), each characterized by a complex amplitude $\alpha_k \in \mathbb{C}$ and a set of dispersion parameters $\bm{\theta}_k = [\varphi_k\iist\tau_k]^{\mathrm{T}}$, where $\tau_k$ denotes the path delay and $\varphi_k$ the angle of arrival (AoA) at the receive array. The according physical channel impulse response (CIR) is modeled as
\vspace*{-1mm}
\begin{align}
    h(\tau,\varphi) = \sum_{k=1}^{K} \alpha_k \ist \delta(\tau - \tau_k) \ist \delta(\varphi - \varphi_k)  \iist  \in \iist \mathbb{C}.
    \label{eq:CIR}\nonumber\\[-7mm]
\end{align}
 For a general planar array with antenna elements at positions $\bm{\rho}_p = [x_p, y_p]^{\mathrm{T}}$ with $p=1,\dots,P$, the response of the array for an incoming wave with AoA $\varphi_k$ is modeled by the array response vector $\bm{a}(\varphi_k) = [a_1(\varphi_k) \iist \cdots \iist a_P(\varphi_k)]^\text{T}$, where the $p$th entry $a_p(\varphi_k) = \exp\!\big(-j \tfrac{2\pi f_{\text{c}}}{c}\ist \bm{\rho}_p^{\mathrm{T}} \bm{u}(\varphi_k)\big)$, $f_{\text{c}}$ is the carrier frequency, $c$ the speed of light, and $\bm{u}(\varphi_k)$ the normalized direction vector. To model antenna-dependent gain and phase deviations, we include complex-valued calibration weights $\bm{w} = [w_1 \iist \cdots \iist w_P]^\text{T} \in \mathbb{C}^{P \times 1}$. The received array signal at antenna $p$ reads
\begin{equation}
   r_p(t)
   = w_p
     \int_{\varphi}  \int_{\tau}a_p(\varphi) \ist s(t-\tau)\ist h(\tau,\varphi)\ist \mathrm{d}\tau \ist \mathrm{d}\varphi
     + \nu_p(t)
   \label{eq:received_signal}
\end{equation}
where $s(t)$ denotes the transmitted signal and $\nu_p(t)$ represents an AWGN process.

\subsection{Discrete Frequency-Domain Signal Model}
\label{subsec:stacked_model}

The signal $r_p(t)$ is  Nyquist filtered, Fourier transformed, and then synchronously and uniformly sampled with frequency spacing $\Delta$ over the bandwidth $B$ to collect a total of $N = B/\Delta$ samples that are arranged in $\bm{y}_p$ as
\vspace*{-1mm}
\begin{align}
    \bm{y}_p
    = \sum_{k=1}^{K} \alpha_k w_p \bm{t}_p(\bm{\theta}_k) + \bm{n}_p \quad \in \mathbb{C}^{N \times 1}
    \label{eq:y_f_antenna}\nonumber\\[-7mm]
\end{align}
where $\bm{t}_p(\bm{\theta}_k) = a_p(\varphi_k)\ist \mathrm{diag}(\bm{s}_{\text{f}})\ist\bm{a}_\tau(\tau_k)$ $\in \mathbb{C}^{N \times 1}$ with $\mathrm{diag}(\cdot)$ denoting a square diagonal matrix with the elements of the vector given as an argument. The vector $\bm{a}_\tau(\tau_k) = \exp\!\big(-j 2\pi \bm{f} \tau_k\big)\in \mathbb{C}^{N\times 1}$ is the temporal response vector, capturing frequency-dependent phase shifts across the bandwidth, with $\bm{f} = \bigl[\, -\tfrac{N}{2}\Delta \;\cdots\; (\tfrac{N}{2}-1 )\Delta \,\bigr]^{\mathrm{T}}$ holding the equally spaced baseband frequency points ($N$ is assumed to be even). The vector $\bm{s}_{\mathrm{f}} = \bigl[\, S(-\tfrac{N}{2}\Delta) \;\cdots\; S((\tfrac{N}{2}-1)\Delta) \,\bigr]^{\mathrm{T}}$ holds the samples of the Fourier spectrum $S(f)$ of $s(t)$, reflecting wideband spectral amplitudes. The measurement noise vector $\bm{n}_p$ denotes a complex circular symmetric Gaussian random vector with covariance matrix $\lambda^{-1}\bm{I}_{N}$, where $\lambda$ is the noise precision.

The signals $\bm{y}_p$ in \eqref{eq:y_f_antenna} are stacked 
$\bm{y}=[\bm{y}_1^{\mathrm{T}}\ist\cdots\ist\bm{y}_P^{\mathrm{T}}]^{\mathrm{T}}\in\mathbb{C}^{PN\times 1}$, which is expressed as
\begin{equation}
    \bm{y} = \bm{D}(\bm{w})\ist\bm{A}(\bm{\theta})\ist\bm{\alpha} + \bm{n}
    \label{eq:y_compact_final}
\end{equation}
where $\bm{\alpha}=[\alpha_1 \ist\cdots\ist\alpha_K]^{\mathrm{T}}$, $\bm{D}(\bm{w})=\mathrm{diag}(\bm{1}_N\otimes\bm{w})$ $\in \mathbb{C}^{PN \times PN}$ and $\bm{A}(\bm{\theta}) =\big[ \bm{a}(\varphi_1) \otimes (\mathrm{diag}(\bm{s}_{\text{f}})\bm{a}_\tau(\tau_1))\ist$ $\cdots \ist\bm{a}(\varphi_K) \otimes (\mathrm{diag}(\bm{s}_{\text{f}})\bm{a}_\tau(\tau_K)) \big]$ $\in \mathbb{C}^{PN \times K}$ with $\otimes$ denoting the Kronecker-product \cite{RichterPhD2005}. Assuming that the measurement noise at different antenna elements is independent and identically distributed, the stacked noise vector $\bm{n} = [\bm{n}_1^{\mathrm{T}} \ist \cdots \ist\bm{n}_P^{\mathrm{T}}]^{\mathrm{T}}$ follows a circularly symmetric complex Gaussian distribution with covariance matrix $\lambda^{-1}\bm{I}_{PN}$.

\section{Problem Formulation and Bayesian Model}
\label{sec:prob_model}

The objective is to jointly estimate the complex multipath amplitudes $\boldsymbol{\alpha}$, dispersion parameters \(\boldsymbol{\theta}\), the number of effective MPCs $K$ as well as the calibration weights of the receive array \(\boldsymbol{w} = [w_1, \ldots, w_P]^{\mathrm{T}}\) and the noise precision parameter \(\lambda\) from the received signal $\bm{y}$ in \eqref{eq:y_compact_final}. 

To estimate the number of MPCs $K$, we extend the sum in \eqref{eq:y_f_antenna} to a fixed (maximum) number of MPCs $K_{\text{max}}\geq K$ by introducing $K_{\text{max}}-K$ ``virtual'' components with amplitude $\alpha_k=0$ and inconsequential parameters $\bm{\theta}_k$ for $k=K\rmv\rmv+\rmv\rmv1,\dots,K_{\text{max}}$. Extending the dimension of the vectors and matrices in \eqref{eq:y_compact_final} accordingly yields a sparse vector $\bm{\alpha}=[\alpha_1\ist\cdots\ist\alpha_K\ist\iist 0\ist\cdots\ist 0]^\T \in \mathbb{C}^{K_{\text{max}}\times 1}$.
We proceed to estimate $K$ indirectly by introducing a sparsity-inducing hierarchical Gamma-Gaussian prior model and obtaining a sparse estimate $\hat{\bm{\alpha}}$ of $\bm{\alpha}$. An estimate $\hat{K}$ of $K$ is obtained as the number of nonzero elements of $\hat{\bm{\alpha}}$.
Following \cite{ShuBucKulPoo:TSP2011,MoePerWitLei:TSP2024}, we model the amplitudes by independent complex Gaussian distributions with component-wise precisions $\gamma_k$, i.e.,
\vspace{-2mm}
\begin{align}
	p(\bm{\alpha}|\bm{\gamma})
	= \prod_{k=1}^{K_{\text{max}}} \mathrm{CN}\ist\big(\alpha_k\ist\big|\ist0,\ist\gamma_k^{-1}\big)
	\label{eq:prior_alpha}\\[-7mm]\nn
\end{align}
where $\bm{\gamma}=[\gamma_1\ist\cdots\ist \gamma_{K_{\text{max}}}]^\T$, and each precision parameter $\gamma_k$ is given by $p(\gamma_k) = \mathrm{Ga}(\gamma_k\ist|\ist\epsilon,\ist\eta)$. Here, $\mathrm{Ga}(\cdot\ist|\ist\epsilon,\ist\eta)$ is a Gamma PDF with shape parameter $\epsilon$ and rate $\eta$. This model is known to encourage many of the amplitudes $\alpha_k$ to be close to zero.
The calibration weights $\bm{w}$ are modeled by independent complex Gaussian PDFs given by\vspace{-1mm}
\begin{align}
	p(\bm{w}) = \prod_{p=1}^{P} \mathrm{CN}\ist \big(w_p\ist\big|\ist\mu_{\text{w},p},\ist\sigma_{\text{w},p}^2 \big)
	\label{eq:prior_w}\\[-7mm]\nn
\end{align}
with means $\mu_{\text{w},p}$ and variances $\sigma_{\text{w},p}^2$. 

The likelihood of the observed signal in \eqref{eq:y_compact_final} is given by
\begin{equation}
	p(\bm{y}|\bm{w},\bm{\alpha},\lambda;\bm{\theta})
	= \mathrm{CN}\!\left(\bm{y}\ist\middle|\ist \bm{D}(\bm{w})\ist\bm{A}(\bm{\theta})\ist\bm{\alpha},\ist \lambda^{-1}\bm{I}\right)
	\label{eq:likelihood}
\end{equation}
 where $\mathrm{CN}(\bm{x};\ist\bm{\mu},\ist\bm{\Sigma})=|\pi \bm{\Sigma}|^{-1} e^{-(\bm{x}-\bm{\mu})^\text{T} \bm{\Sigma}^{-1}(\bm{x}-\bm{\mu})}$ denotes the PDF of multivariate complex Gaussian random variable $\bm{x}$ with mean $\bm{\mu}$ and covariance $\bm{\Sigma}$. The noise precision $\lambda$ is modeled as an independent Gamma random PDF $p(\lambda) = \mathrm{Ga}(\lambda\ist|\ist a,\ist b)$ providing conjugacy for the Gaussian likelihood and enabling joint estimation of the noise variance.

Based on this introduced model, the according joint posterior PDF is proportional to
\begin{align}
    &p(\bm{w},\bm{\alpha},\bm{\gamma},\lambda\ist|\ist\bm{y};\ist\bm{\theta})\nn\\
    & \hspace{12mm} \propto
    p(\bm{y}|\bm{w},\bm{\alpha},\lambda;\bm{\theta})\ist
    p(\bm{w})\ist
    p(\bm{\alpha}|\bm{\gamma})\ist
    p(\bm{\gamma})\ist
    p(\lambda)\ist.
    \label{eq:joint}
\end{align}

Since the computation of the exact posterior PDF is intractable, we resort to a variational inference framework using a structured mean-field approximation \cite{TziLikGal:SPM2008}. The resulting variational EM algorithm alternates between (i) updating proxy posterior distributions of $\bm{w}$, $\bm{\alpha}$, and $\lambda$ in the E-step, and (ii) applying fast updates for $\bm{\gamma}$ and the deterministic parameters $\bm{\theta}$ in the M-step.

\section{Variational Bayesian Inference}

\subsection{Derivation of proxy PDFs $  q_{\bm{w}}$, $q_{\bm{\alpha}}$, $q_\lambda$, and $q_{\gamma,k}$}

We aim to obtain point estimates $\hat{\bm{\theta}}$ of $\bm{\theta}$ while approximating the posterior PDF of all other involved parameters by a ``simpler'' factorized proxy PDF given by
\begin{align}
	q(\bm{w},\bm{\alpha},\bm{\gamma},\lambda; \hat{\bm{\theta}})&= q_{\bm{w}}(\bm{w})q_{\bm{\alpha}}(\bm{\alpha})\ist q_{\lambda}(\lambda) \prod_{k=1}^{K_{\text{max}}}q_{\gamma,k}(\gamma_{k})	
    \, .
\end{align}
This factorization of proxy PDFs is referred to as the mean-field approximation.
Given the current estimate $\hat{\bm{\theta}}$ of $\bm{\theta}$, the proxy PDFs are obtained by minimizing the Kullback--Leibler (KL) divergence  between the true and proxy PDF in the expectation step, which is equivalent to maximizing the evidence lower bound (ELBO), i.e.,
\vspace{-1mm}
\begin{align}
    q &= \arg\max_{q \in \mathcal{Q}} \mathcal{L}\nonumber\\[-7mm]
\end{align}
where $\mathcal{Q}$ denotes the family of proxy distributions over $\{\bm{w},\bm{\alpha},\bm{\gamma},\lambda\}$ and the ELBO $\mathcal{L}$ is defined as
\begin{align}
    \hspace*{-2mm}\mathcal{L}(q;\bm{\theta})  \hspace*{-0.2mm}&=\hspace*{-0.2mm}\big< \ln p(\bm{w},\bm{\alpha},\bm{\gamma},\lambda,\bm{y};\bm{\theta})
    \hspace*{-0.2mm}-\hspace*{-0.2mm}\ln q(\bm{w},\bm{\alpha},\bm{\gamma},\lambda;\bm{\theta}) \big>_q 
\end{align}
where $\langle \cdot \rangle_q$ denotes the expectation with respect to $q(\bm{w},\bm{\alpha},\bm{\gamma},\lambda)$. This results in the following consistency equations for respective $q_j$ that are iteratively executed \cite{TziLikGal:SPM2008}
\begin{equation}
	q^\star_j \propto \exp \big<\ln p(\bm{w},\bm{\alpha},\bm{\gamma},\lambda,\bm{y};\hat{\bm{\theta}})\big>_{\bar{q}_{j}}
    \label{eq:proxy_pdf}
\end{equation}
where $\bar{q}_{j}=\prod_{q_i \in \mathcal{Q}\backslash q_j} q_i$ denotes the product of all factors of the joint proxy $q$ except $q_j$.

\subsubsection*{Consistency equation for $q^\star_{\bm{w}}$} For $\bm{w}$, we obtain
\begin{align}
    q^\star_{\bm{w}} &= \text{CN}(\bm{w}|\hat{\bm{w}},\hat{\bm{\Sigma}}_{\text{w}})
    \label{eq:q-w}
\end{align}
where $\hat{\bm{w}}=[\hat{w}_1\ist\cdots\ist\hat{w}_P]^\T$ and $\hat{\bm{\Sigma}}_{\text{w}} = \text{diag}(\hat{\bm{\sigma}}^2_{\text{w}})$ with $\hat{\bm{\sigma}}^2_{\text{w}}=[\hat{\sigma}^2_{\text{w},1} \ist\cdots\ist \hat{\sigma}^2_{\text{w},P}]^\T$.
That is, $q^\star_{\bm{w}}$ is a product of independent complex Gaussian PDFs with mean $\hat{w}_p$ and variance $\hat{\sigma}_{\text{w},p}^2$, respectively, which are given by (see the appendix)
\begin{align}
\hat{\sigma}_{\text{w},p}^2 &= \big(\hat{\lambda}[\hat{\bm{\alpha}}^{\mathrm{H}}\hat{\bm{T}}^{\mathrm{H}}_{p}\hat{\bm{T}}_{p}\hat{\bm{\alpha}} + \text{tr}(\hat{\bm{T}}^{\mathrm{H}}_{p}\hat{\bm{T}}_{p}\hat{\bm{\Sigma}}_{\alpha})] +\sigma_{\text{w},p}^{-2}\big)^{-1} \nonumber \\
	\hat{w}_p &= \hat{\sigma}_{\text{w},p}^2\big(\hat{\lambda} \hat{\bm{\alpha}}^{\mathrm{H}}\hat{\bm{T}}^{\mathrm{H}}_{p}\bm{y}_p  + \sigma_{\text{w},p}^{-2} \ist \mu_{\text{w},p}\big)
    \label{eq:q_w}
\end{align}
where $\hat{\bm{T}}_p \triangleq \bm{T}_p(\hat{\bm{\theta}})$,  $\bm{T}_p(\bm{\theta}) = [\bm{t}_p(\bm{\theta}_1) \iist \cdots \iist \bm{t}_p(\bm{\theta}_K)]^{\text{T}}$ $\in \mathbb{C}^{N \times K}$, and $\hat{\bm{\alpha}}$ and $\hat{\bm{\Sigma}}_{\alpha}$ are the mean and covariance of $q_{\bm{\alpha}}$, respectively, see \eqref{eq:q_alpha}. 
Note that the prior mean $\mu_{\text{w},p}$ may stem from an a priori calibration of the antenna system, whereas the prior variance $\sigma_{\text{w},p}^{2}$ quantifies the confidence in these apriori values, with smaller variances corresponding to higher trust.

\subsubsection*{Consistency equation for $q^\star_{\bm{\alpha}}$} For $\bm{\alpha}$ we obtain
\begin{align}
    q^\star_{\bm{\alpha}} &= \text{CN}(\bm{\alpha}\ist|\ist\hat{\bm{\alpha}},\ist  \hat{\bm{\Sigma}}_\alpha) 
    \label{eq:q-alpha}
\end{align}
i.e., a complex Gaussian PDF with mean $\hat{\bm{\alpha}}$ and variance $\hat{\bm{\Sigma}}_\alpha$ given by
\begin{align}
\hat{\bm{\Sigma}}_{\alpha} &= \big(\hat{\lambda}\ist\hat{\bm{A}}^{\mathrm{H}} (|\hat{\bm{D}}_{\text{w}}|^2+\hat{\bm{D}}_{\sigma})\hat{\bm{A}} +\text{diag}(\hat{\bm{\gamma}})\big)^{-1}\ist \nonumber \\
\hat{\bm{\alpha}} &= \hat{\lambda}\ist\hat{\bm{\Sigma}}_{\alpha}\hat{\bm{A}}^{\mathrm{H}}\hat{\bm{D}}_{\text{w}} ^{\mathrm{H}}\bm{y}
\label{eq:q_alpha}
\end{align}
where $\hat{\bm{A}} \triangleq \bm{A}(\hat{\bm{\theta}})$, $\hat{\bm{D}}_{\text{w}} \triangleq \bm{D}(\hat{\bm{w}})$, and $\hat{\bm{D}}_{\sigma} \triangleq \bm{D}(\hat{\bm{\sigma}}_{\text{w}}^2)$.

\subsubsection*{Consistency equation for $q^\star_{\lambda}$} For $\lambda$, we obtain
\begin{align}
    q^\star_\lambda= \text{Ga}(\lambda\ist |\ist a+NP,\ist b + \rho) 
\label{eq:q-lambda} 
\end{align}
with $\rho = \big\| \bm{y} - \hat{\bm{D}}_{\text{w}} \hat{\bm{A}} \hat{\bm{\alpha}} \big\|^2  + \mathrm{tr} \big( (|\hat{\bm{D}}_{\text{w}}|^2 + \hat{\bm{D}}_{\sigma}) \hat{\bm{A}}\hat{\bm{\Sigma}}_\alpha \hat{\bm{A}}^\mathrm{H} ) + \hat{\bm{\alpha}}^\mathrm{H}\hat{\bm{A}}^\mathrm{H}\hat{\bm{D}}_{\sigma}\hat{\bm{A}}\hat{\bm{\alpha}} $. The optimal factors depend on $q_\lambda$ only via its mean, i.e., $\hat{\lambda}=\big<\lambda\big>_{q_\lambda}$, given by
\begin{align}
	\hat{\lambda} &= \frac{a+NP}{b +\rho}\ist.
	\label{eq:update_lambda}
\end{align}

\subsubsection*{Consistency equation for $q^\star_{\gamma,k}$} For $\gamma_k$, we obtain
\begin{align}
q_{\gamma,k} &= \text{Ga}(\gamma_{k}\ist |\ist \epsilon+1,\ist \eta + \hat{\Sigma}_{\alpha,kk} + |\hat{\alpha}_k|^2)
\label{eq:q-gamma}
\end{align}
where $\hat{\Sigma}_{\alpha,kk}$ is the $k$th element of the main diagonal of $\hat{\bm{\Sigma}}_{\alpha}$ and $\hat{\alpha}_k$ is the $k$th element of $\hat{\bm{\alpha}}$.
The optimal factors depend on $q_{\gamma,k}$ only via its mean, i.e., $\hat{\gamma}_k=\big<\gamma_k\big>_{q_{\gamma,k}}$, given by
\begin{equation}
    \hat{\gamma}_k = \frac{\epsilon+1}{\eta + \hat{\Sigma}_{\alpha,kk} + |\hat{\alpha}_{k}|^2}
    \ist .
\end{equation}


\subsection{Fast updates of $q_{\bm{\alpha}}$, $q_{\gamma,k}$ and $\hat{\bm{\theta}}_k$ for $k=1,\dots,K_{\text{max}}$}
There exists a strong interdependence between the variables $\bm{\alpha}$, $\gamma_k$, and $\bm{\theta}_k$, $k=1,\dots,K_{\text{max}}$, which often results in slow convergence of the iterative estimation procedure described above. Thus, we propose a joint update of $q_{\bm{\alpha}}$, $q_{\gamma,k}$, and $\hat{\bm{\theta}}_k$ for $k = 1, \ldots, K_{\text{max}}$ instead.

Repeatedly alternating between updates of $q_{\bm{\alpha}}$ and $q_{\gamma,k}$ leads to a first-order recursive sequence of estimates $\hat{\gamma}_k$ \cite{ShuBucKulPoo:TSP2011,moederlFusion2025:multi-dictionary-SBL,MoeLeiFlePerWit:TSP2025}. If a stationary point of this sequence exists, it can be determined in closed form, enabling a fast update of $\hat{\gamma}_k$.
This fast update is equivalent to maximizing the marginal likelihood
\begin{equation} \label{eq:sbl-marginal-likelihood}
    L(\bm{\theta},\bm{\gamma};\hat{\lambda},\hat{\bm{w}}) = \int p(\bm{y}|\hat{\bm{w}},\bm{\alpha},\hat{\lambda};\bm{\theta})\ist
    p(\bm{\alpha}|\bm{\gamma})\ist
    \tilde{p}(\bm{\gamma}) \,\mathrm{d}\bm{\alpha}
\end{equation}
with respect to $\gamma_k$ while keeping the remaining elements of $\bm{\gamma}$ fixed, where $\tilde{p}(\bm{\gamma})=\prod_{k=1}^{K_{\text{max}}}\tilde{p}(\gamma_k)$ is an ``equivalent prior'' \cite[Corollary~2]{MoeLeiFlePerWit:TSP2025}.
    \footnote{Specifically, $p(\gamma_k)=\text{Ga}(\gamma_k|\epsilon,\eta)$ with $\epsilon=\eta=0$ yields the equivalent prior $\tilde{p}(\gamma_k)\propto 1$.}
For any $k=1,\dots,K_{\text{max}}$, \eqref{eq:sbl-marginal-likelihood}, the dependence of $L(\bm{\theta},\bm{\gamma};\hat{\lambda},\hat{\bm{w}})$ on $(\bm{\theta}_k,\gamma_k)$, can be expressed as \cite{moederlFusion2025:multi-dictionary-SBL}
\begin{align} \label{eq:partial-likelihood2}
        \ell_k(\gamma_k, \bm{\theta}_k)
        = \frac{|\mu_k(\bm{\theta}_k)|^2 / s_k(\bm{\theta}_k)}{1 + \gamma_k s_k(\bm{\theta}_k)}
        + \log \frac{\gamma_k s_k(\bm{\theta}_k)}{1 + \gamma_k s_k(\bm{\theta}_k)}
\end{align}
with
\begin{align}
    s_k(\bm{\theta}_k) &= \!\big(\hat{\lambda}\, \bm{d}_k^{\mathrm{H}}\bm{d}_k
    - \hat{\lambda}^2 \bm{d}_k^{\mathrm{H}}\hat{\bm{D}}_{\bar{k}}
      \hat{\bm{\Sigma}}_{\alpha,\bar{k}}\hat{\bm{D}}_{\bar{k}}^{\mathrm{H}}\bm{d}_k\big)^{-1} \nonumber\\
   \mu_k(\bm{\theta}_k) &= \hat{\lambda}s_k\, \bm{d}_k^{\mathrm{H}}\bm{y}
   - \hat{\lambda}^2 s_k\, \bm{d}_k^{\mathrm{H}}\hat{\bm{D}}_{\bar{k}}
   \hat{\bm{\Sigma}}_{\alpha,\bar{k}}\hat{\bm{D}}_{\bar{k}}^{\mathrm{H}}\bm{y}
\end{align}
where $\bm{d}_k = \hat{\bm{D}}_{\text{w}}\bm{A}_k(\bm{\theta}_k)$ implicitly depends on $\bm{\theta}_k$ with $\bm{A}_k \in \mathbb{C}^{PN \times 1}$ denoting the $k$th column of $\bm{A}(\bm{\theta})$. Here, $\hat{\bm{D}}_{\bar{k}} = \hat{\bm{D}}_{\text{w}}\hat{\bm{A}}_{\bar{k}}$ with $\hat{\bm{A}}_{\bar{k}}$ denoting the matrices obtained by removing the $k$th column from $\hat{\bm{A}}$. The same applies to $\hat{\bm{\Sigma}}_{\alpha,\bar{k}}$.
The cost function in \eqref{eq:partial-likelihood2} can be jointly maximized with respect to $\gamma_k$ and $\bm{\theta}_k$ by \cite{moederlFusion2025:multi-dictionary-SBL}
\begin{equation}
    \hat{\bm{\theta}}_k
    = \arg\max_{\bm{\theta}_k} \frac{|\mu_k(\bm{\theta}_k)|^2}{s_k(\bm{\theta}_k)}
    \label{eq:CostParameters}
    \, 
\end{equation}
and
\begin{equation}
\hat{\gamma}_k =
\begin{cases}
(|\mu_k(\hat{\bm{\theta}}_k)|^2 - s_k(\hat{\bm{\theta}}_k))^{-1}, & \text{if }\frac{|\mu_k(\hat{\bm{\theta}}_k)|^2}{s_k(\hat{\bm{\theta}}_k)} > \chi, \\[4pt]
\infty, & \text{otherwise}.
\end{cases}
\label{eq:GammaUpdate}
\end{equation}
Here, $\chi \geq 1$ denotes the pruning threshold used to suppress spurious component detections, as discussed in \cite{LeiGreFleWit:CSSC2020}.
Once updated estimates $\hat{\gamma}_k$ and $\hat{\bm{\theta}}_k$ are obtained, we update $q_{\bm{\alpha}}$ using \eqref{eq:q_alpha}.

We iteratively perform joint updates of the channel parameters $q_{\bm{\alpha}}$, $q_{\gamma_k}$ and $\hat{\bm{\theta}}_k$ for $k=1,\dots,K_{\text{max}}$, followed by updating the noise precision $\hat{\lambda}$ using \eqref{eq:update_lambda} and updates of the calibration weights $q_{\bm{w}}$ using \eqref{eq:q_w} until a convergence criterion is fulfilled or a maximum number of iterations is reached.

\section{Algorithm Implementation}
We have implemented Algorithm~\ref{alg:parameter_estimation}, which includes the update formulas of the proxy PDFs $q_{\bm{w}}$, $q_{\bm{\alpha}}$ and $q_{\lambda}$, as well as the fast updates for the component-wise precisions $\bm{\gamma}$ and the component parameters $\bm{\theta}$. The algorithm keeps track of the nonzero components of our model. We start with an empty model using Jeffrey's priors ($a = b =\epsilon =\eta$ = 0) for the variables $\bm{\gamma}$ and $\lambda$ and a non-informative prior on the weights ($\sigma_{\text{w},p}^2=100$), assuming no antenna element imperfections ($\mu_{\text{w},p}=1$). Thus, we initialize $\hat{\bm{D}}_{\text{w}} = \mathrm{diag}(\bm{1}_{PN})$ and $\hat{\bm{D}}_{\sigma} = \mathrm{diag}(\bm{0}_{PN})$, reflecting the absence of any influence from weight imperfections before the first calibration weight update. The noise precision is initialized as $PN/||\bm{y}||^2$. Our algorithm iteratively detects, refines, and prunes MPCs from the received signal.
In each iteration, we first search for a new MPC to add to the model by maximizing the objective function \eqref{eq:CostParameters} using Matlab's \verb|fminsearch| function. To aid the search, we initialize the numeric optimization with a coarse estimate obtained by finding the maximum of the beamformer $|\bm{d}_k^\H(\bm{\theta}_k)\bm{y}_{\text{res}}|^2$ evaluated on a predefined parameter grid $\Theta_{\text{grid}}$, where $\bm{y}_{\text{res}}=\bm{y}-\hat{\bm{D}}_{\text{w}}\hat{\bm{A}}\hat{\bm{\alpha}}$.
 If $\frac{|\mu_k(\hat{\bm{\theta}}_k)|^2}{s_k(\hat{\bm{\theta}}_k)} $ of the new candidate exceeds the pruning threshold $\chi$, it is added to the model. After detection, the algorithm updates the proxy PDF for the calibration weights $q_{\bm{w}}$. Each existing component is then refined, using the fast update formulas for $\hat{\bm{\theta}}_k$ and $\hat{\gamma}_k$, potentially pruning weak components. This process of detection, component and parameter updates, and pruning continues until the estimated number of components $\hat{K}$, their parameters $\hat{\bm{\theta}}$, the calibration weights $\hat{\bm{w}}$ and the noise precision $\hat{\lambda}$ converge.
 
\begin{algorithm}
\caption{FVSBL Algorithm with Self-Calibration}
 \label{alg:parameter_estimation}  
\textbf{Inputs:} $\bm{y}$, $\chi$, and $\Theta_{\mathrm{grid}}$ \\
\textbf{Outputs:}  $\hat{K}$, $\hat{\bm{\theta}}$, $\hat{\bm{w}}$, and $\hat{\lambda}$
\begin{algorithmic}[1]
\State Init: $\hat{K}=0$, $\hat{\bm{\theta}}\gets [\ist],\hat{\bm{\gamma}}\gets [\ist]$,  $\hat{\lambda}\gets \frac{PN}{\|\bm{y}\|^2}$, $\sigma_{\text{w},p}^2=100$, $\mu_{\text{w},p}=1$ for $p=1,\dots, P, \, \hat{\bm{D}}_{\text{w}}=\mathrm{diag}(\bm{1}_{PN})$, and \quad $\hat{\bm{D}}_{\sigma}=\mathrm{diag}(\bm{0}_{PN})$.
\Repeat
  
    \State \textbf{Detection Phase:}
    \State $k\leftarrow \hat{K}+1$.
    \State $\bm{y}_{\text{res}} \gets \bm{y} - \hat{\bm{D}}_{\text{w}}\hat{\bm{A}}\hat{\bm{\alpha}}$ if $\hat{K}>0$ else $\bm{y}_{\text{res}}\gets \bm{y}$.
    \State $\hat{\bm{\theta}}_{\text{init}} \gets \arg\max_{\bm{\theta}_k\in\Theta_{\mathrm{grid}}} \left| \bm{d}_k^\H(\bm{\theta}_k) \bm{y}_{\text{res}} \right|^2$.
    \State $\hat{\bm{\theta}}_k \gets \arg\max_{\bm{\theta}_k} \frac{|\mu_k(\bm{\theta}_k)|^2}{s_k(\bm{\theta}_k)}$ (search initialized at $\hat{\bm{\theta}}_{\text{init}}$).
    \If{$\frac{|\mu_k(\hat{\bm{\theta}}_k)|^2}{s_k(\hat{\bm{\theta}}_k)} > \chi$}
        \State Add component: $\hat{K}\gets \hat{K}+1$, $\hat{\bm{\theta}} \gets \big[\hat{\bm{\theta}}^\T \ist\iist \hat{\bm{\theta}}_k^\T\big]^\T$,
        \State $\hat{\bm{\gamma}} \gets \big[\hat{\bm{\gamma}}^\T\ist\iist \hat{\gamma}_k]^\T$ with $\hat{\gamma}_k$ calculated by \eqref{eq:GammaUpdate}.
        \State \textbf{Amplitude and Noise Update:}
    \State Update $q_{\bm{\alpha}}, q_{\lambda}$ according to (\ref{eq:q_alpha}), (\ref{eq:update_lambda}).
    \EndIf
    \State \textbf{Calibration Weight Update:}
    \State Update $q_{\bm{w}}$ according to (\ref{eq:q_w}).
    \State \textbf{Component Update:}
    \For{$k = 1$ to $\hat{K}$}
         \State Refine $\hat{\bm{\theta}}_k$ by maximization of (\ref{eq:CostParameters}).
         \If{$\frac{|\mu_k(\hat{\bm{\theta}}_k)|^2}{s_k(\hat{\bm{\theta}}_k)} > \chi$}
            \State Update $\hat{\gamma}_k$ using \eqref{eq:GammaUpdate}.
         \Else
            \State Remove component: $\hat{K} \gets \hat{K}-1$, and
            \State remove $\hat{\bm{\theta}}_k$ from $\hat{\bm{\theta}}$ and $\hat{\gamma}_k$ from $\hat{\bm{\gamma}}$.
         \EndIf
    \EndFor
    \State \textbf{Amplitude and Noise Update:}
    \State Update $q_{\bm{\alpha}}, q_{\lambda}$ according (\ref{eq:q_alpha}), (\ref{eq:update_lambda}).
\Until{convergence criterion is met}.
\end{algorithmic}
\end{algorithm} 

\section{Statistical Evaluation}
We have applied our algorithm to synthetic data generated for a uniform linear array with $P=4$ antenna elements, spaced with $\frac{c}{2f_{\text{c}}}$ (i.e., half-wavelength), and a bandwidth of $2$ GHz.
We compare our algorithm to a VSBL channel estimation algorithm without calibration. This algorithm is obtained by assuming a very low prior variance of the calibration weights $\sigma^2_{\text{w},p}=10^{-8}$ in the update step of the weights (Line 15 in Algorithm~
\ref{alg:parameter_estimation}), effectively disabling their estimation. As further benchmark we also include an ``optimal'' variant which is given the true calibration weights instead of estimating them.
The synthetic measurement data is generated from  $K=3$ well-separated MPCs with high component SNRs, i.e., $40$~dB, $38$~dB and $35$~dB, demonstrating that integrated calibration improves channel parameter estimation when antenna element imperfections are present.

For evaluation, we map the estimated delays $\hat{\bm{\tau}}$ to equivalent distances $\hat{\bm{\tau}}_d = c \hat{\bm{\tau}}$.
We evaluate the angle and delay estimation errors separately. We use the OSPA metric \cite{SchVoVo:TSP2008} to jointly account for estimation errors and cardinality errors.
The cutoff parameters are set to $0.1$ m for distances and $50^\circ$ for angles. We also investigate the estimated number of components $\hat{K}$ and the convergence time $t_{\mathrm{conv}}$,
\footnote{Using MATLAB R2023b on an AMD Ryzen 7 PRO 5875U processor.}
comparing scenarios with and without self-calibration under calibration mismatches. The simulated weights are drawn from a complex normal distribution $\bm{w} \sim \text{CN}(1, \sigma_{\text{w,sim}}^2)$. For each value of $\sigma_{\text{w,sim}}^2$, the metrics are averaged over 100 simulation runs.


In Fig.~\ref{fig:OSPAs}, the OSPA metrics for the estimated channel parameters ($\hat{\bm{\tau}}_d$, $\hat{\bm{\varphi}}$) are plotted against the simulation standard deviations of the weights $\sigma_{\text{w,sim}}$. The OSPA values are significantly higher when the calibration weight update is excluded, but only for simulation standard deviations above $3 \times 10^{-2}$. For simulation standard deviations below $3 \times 10^{-2}$, the weight deviations become negligible. Comparing the optimal estimation results for distances and angles, we see that the calibration-related errors can be eliminated for the OSPA $\tau_d$ and substantially reduced for the OSPA $\varphi$. This indicates that the gains of the deviation weights are estimated more precisely than the phases. In Fig.~\ref{fig:K_hats}, the number of estimated components $\hat{K}$  is plotted against $\sigma_{\text{w,sim}}$. Numerous cardinality errors occur due to the additional estimation of spurious MPCs when calibration is not applied. We identified this as the primary reason for the increase in the OSPA values shown in Fig.~\ref{fig:OSPAs}. Finally, Fig.~\ref{fig:estimation_time} shows the convergence time $t_{\mathrm{conv}}$ of the proposed algorithm compared to the non-calibration scenario. The calibration variant maintains a relatively constant convergence speed, whereas $t_{\mathrm{conv}}$ increases in the no-calibration scenario once calibration deviations become relevant.

\section{Conclusion} 

In this work, we present a novel Bayesian method for self-calibration of antenna element phase and gain imperfections within a VSBL-based wideband channel estimation framework. The complex calibration weights of the antenna elements are incorporated as a diagonal matrix into a wideband array signal model, enabling the joint probabilistic estimation of channel parameters and calibration weights. By leveraging a fast variant of VSBL, we derive closed-form update equations for the proxy PDFs and model parameters and evaluate the proposed algorithm using synthetic data.

The statistical evaluation reveals a substantial improvement in the OSPA metrics for the estimated channel parameters $\hat{\tau}_d$ and $\hat{\varphi}$ when calibration is included, particularly for weight standard deviations exceeding $3\times10^{-2}$. In this regime, the proposed algorithm also outperforms its non-calibration variant with regard to convergence speed. In practical phased-array frontends, weight standard deviations are typically around $10^{-1}$, and the results demonstrate that the proposed method performs especially well in this realistic range, highlighting its potential for modern 6G systems. Future work will focus on validating the approach using real measurement data and extending the algorithm to more sophisticated models \cite{MoePerWitLei:TSP2024}, capable of capturing more complex antenna impairments (such as mutual antenna coupling).

\begin{figure}[H]
    \centering
    \setlength{\figureheight}{3.5cm}
    \setlength{\figurewidth}{\columnwidth}
    \input{pgf/ospa_tau_plot}
    \vspace*{1mm}
    \makebox[0pt]{(a)}
    \setlength{\figureheight}{3.5cm}
    \setlength{\figurewidth}{\columnwidth}
    \input{pgf/ospa_phi_plot}
    \vspace*{-2mm}
    \makebox[0pt]{(b)}
    \caption{OSPA metrics for estimated channel parameters $\hat{\bm{\tau}}_d$ (a) and $\hat{\bm{\varphi}}$ (b) versus standard deviations of simulation weights $\sigma_{\text{w,sim}}$, comparing the calibration, no-calibration and optimal scenarios. }
    \label{fig:OSPAs}
\end{figure}

\begin{figure}[H]
    \centering
    \setlength{\figureheight}{3.5cm}
    \setlength{\figurewidth}{\columnwidth}
    \input{pgf/K_hats}
    \caption{Number of estimated components $\hat{K}$ versus standard deviations of simulation weights $\sigma_{\text{w,sim}}$, comparing the calibration and no-calibration scenarios.}
    \label{fig:K_hats}
\end{figure}

\begin{figure}[H]
    \centering
    \setlength{\figureheight}{3.5cm}
    \setlength{\figurewidth}{\columnwidth}
    \input{pgf/mean_estimation_times}
    \caption{Convergence time $t_{conv}$ versus standard deviations of simulation weights $\sigma_{\text{w,sim}}$, comparing the calibration and no-calibration scenarios.}
    \label{fig:estimation_time}
\end{figure}


\section*{Appendix: Derivation of $q_{\bm{w}}$} 

The update equation for the proxy PDF $q_{\bm{w}}$ is expressed as
\begin{align}
    \ln q_{\bm{w}}(\bm{w}) 
    &\overset{e}{\propto} 
    \underbrace{\big\langle \ln p(\bm{y}|\bm{w},\bm{\alpha},\lambda;\hat{\bm{\theta}})\big\rangle_{q_{\bm{\alpha}}q_{\lambda}}}_{\text{I}}
    + \underbrace{\ln p(\bm{w})}_{\text{II}}
\end{align}
where the first term (I) represents the expected log-likelihood under the proxy distributions $q_{\bm{\alpha}}$ and $q_{\lambda}$, and the second term (II) denotes the log-prior of $\bm{w}$.  

For a single antenna element $p$, the likelihood can be written as
\begin{align}
    \ln p(\bm{y}_p|w_p,\bm{\alpha},\lambda;\hat{\bm{\theta}})
    &\overset{e}{\propto}  -\lambda\ist\big\| \bm{y}_p - w_p\hat{\bm{T}}_p\bm{\alpha} \big\|^2
\end{align}
with $\bm{y}_p \in \mathbb{C}^{N\times 1}$ and $\hat{\bm{T}}_p \triangleq \bm{T}_p(\hat{\bm{\theta}})$.  
Expanding the square and neglecting all terms independent of $w_p$ yields
\begin{align}
    - 2\Re\{w_p^{\ast}\ist\bm{\alpha}^{\mathrm{H}}\hat{\bm{T}}_p^{\mathrm{H}} \bm{y}_p\}
    + |w_p|^2\bm{\alpha}^{\mathrm{H}}\hat{\bm{T}}^{\mathrm{H}}_p\hat{\bm{T}}_p \bm{\alpha}.
    \label{eq:quadratic_expansion}
\end{align}

The expectation in term~(I) is taken with respect to the current proxy posteriors $q_{\bm{\alpha}}(\bm{\alpha})$ and $q_{\lambda}(\lambda)$, i.e., $\big\langle{\bm{\alpha}} \big\rangle_{q_{\bm{\alpha}}} = \hat{\bm{\alpha}}$, $\big\langle{\lambda} \big\rangle_{q_{\lambda}} = \hat{\lambda}$, $\big\langle \bm{\alpha}^{\mathrm{H}}\hat{\bm{T}}^{\mathrm{H}}_p\hat{\bm{T}}_p\bm{\alpha} \big\rangle_{q_{\bm{\alpha}}} = \hat{\bm{\alpha}}^{\mathrm{H}}\hat{\bm{T}}^{\mathrm{H}}_p\hat{\bm{T}}_p\hat{\bm{\alpha}} + \text{tr}(\hat{\bm{T}}^{\mathrm{H}}_p\hat{\bm{T}}_p\hat{\bm{\Sigma}}_\alpha)\vspace{0.5mm}$ \cite[Eq.~(378)]{Petersen2008}. Applying these expectations to~\eqref{eq:quadratic_expansion} yields
\begin{align}
\text{I}:& \ist\hat{\lambda} \ist\Big(
   2\Re\{w_p^{\ast}\ist \hat{\bm{\alpha}}^{\mathrm{H}} \hat{\bm{T}}_p^{\mathrm{H}} \bm{y}_p\}
    \nn \\
    & - |w_p|^2 \ist\!\big(
       \hat{\bm{\alpha}}^{\mathrm{H}}\hat{\bm{T}}_p^{\mathrm{H}}\hat{\bm{T}}_p \hat{\bm{\alpha}} + \operatorname{tr}(\hat{\bm{T}}_p^{\mathrm{H}}\hat{\bm{T}}_p\hat{\bm{\Sigma}}_\alpha)
     \big)
   \Big).
\end{align}

The prior term (II), $\ln p(w_p) = \ln \text{CN}(w_p\ist|\ist \mu_{\text{w},p},\sigma_{\text{w},p}^2)$, contributes
\begin{align}
    \ln p(w_p) \overset{e}{\propto}  -\sigma_{\text{w},p}^{-2}|w_p-\mu_{\text{w},p}|^2,
\end{align}
which expands to
\begin{equation}
\text{II}: \ln p(w_p) \overset{e}{\propto} -\sigma_{\text{w},p}^{-2} |w_p|^2+ 
2\Re\{\sigma_{\text{w},p}^{-2}w_p^*\mu_{\text{w},p}\}.
\end{equation}

Combining the likelihood (I) and prior (II) terms and collecting coefficients leads to
\begin{align}
    \ln q_{w_p}\overset{e}\propto -A_p |w_p|^2 + 2\Re\{w_p^{\ast}B_p\}.
    \label{eq:quadratic_form}
\end{align}
Here, the coefficients are defined as
\begin{align}
    A_p &= \hat{\lambda} \ist
   \big(
       \hat{\bm{\alpha}}^{\mathrm{H}}\hat{\bm{T}}_p^{\mathrm{H}}\hat{\bm{T}}_p \hat{\bm{\alpha}} + \operatorname{tr}(\hat{\bm{T}}_p^{\mathrm{H}}\hat{\bm{T}}_p\hat{\bm{\Sigma}}_\alpha)
     \big)+ \sigma_{\text{w},p}^{-2} \\
    B_p &= 
    \hat{\lambda}  \ist
    \hat{\bm{\alpha}}^{\mathrm{H}} \hat{\bm{T}}_p^{\mathrm{H}} \bm{y}_p
    + \sigma_{\text{w},p}^{-2}\mu_{\text{w},p}.
\end{align}

Completing the square in~\eqref{eq:quadratic_form} gives
\begin{align}
 -A_p\!\left(|w_p|^2
 - 2\Re\{w_p^{\ast}\tfrac{B_p}{A_p}\}\right) 
    &= -A_p \big|w_p - \tfrac{B_p}{A_p}\big|^2 \;+\; \tfrac{|B_p|^2}{A_p}.
\end{align}
The last term acts only as a normalization constant and does not affect the distribution shape.  
Hence, $q_{\bm{w}}(\bm{w})=\prod_{p=1}^{P}\text{CN}(w_p \ist\big|\ist \hat{w}_{p},\ist \hat{\sigma}_{\text{w},p}^2)$ is a product of independent complex Gaussian distributions with parameters
\begin{align}
     \hat{w}_{p} &=\hat{\sigma}_{\text{w}, p}^2 B_p,
     &
     \hat{\sigma}_{\text{w},p}^2 &=\frac{1}{A_p}
\end{align}
for $p=1,\dots,P$.

\bibliographystyle{IEEEtran}
\bibliography{IEEEabrv,refs_library.bib}

\end{document}

%% file: pgf/ospa_tau_plot.tex
\begin{tikzpicture}
\begin{axis}[
    width=\figurewidth,
    height=\figureheight,
    grid=both,
    xmode=log,
    ymode=log,
    xmin=9e-3, xmax=1e0,
    ylabel= {OSPA $\tau_d$ (m)},
    legend style={font=\footnotesize, at={(0.05,0.95)}, anchor=north west},
]

\addplot [mark=square*, blue, thick] 
    table [col sep=comma, x=sigma_sim2, y=OSPA_tau_cal] {pgf/OSPA_tau.csv};
\addlegendentry{cal}

\addplot [mark=o, red, thick, dashed, mark options={style=solid}] 
    table [col sep=comma, x=sigma_sim2, y=OSPA_tau_nocal] {pgf/OSPA_tau.csv};
\addlegendentry{no cal}

\addplot [green, thick, dashed, mark options={style=solid}] 
    table [col sep=comma, x=sigma_sim2, y=OSPA_tau_optcal] {pgf/OSPA_tau.csv};
\addlegendentry{optimal}

\end{axis}
\end{tikzpicture}

%% file: pgf/ospa_phi_plot.tex
\begin{tikzpicture}
\begin{axis}[
    width=\figurewidth,
    height=\figureheight,
    grid=both,
    xmode=log,
    ymode=log,
     xmin=9e-3, xmax=1e0,
    xlabel={$\sigma_{\text{w,sim}}$},
    ylabel={OSPA $\varphi$ ($^\circ$)},
    legend style={font=\footnotesize, at={(0.05,0.95)}, anchor=north west},
]

\addplot [mark=square*, blue, thick] 
    table [col sep=comma, x=sigma_sim2, y=OSPA_phi_cal] {pgf/OSPA_phi.csv};
\addlegendentry{cal}

\addplot [mark=o, red, thick, dashed, mark options={style=solid}] 
    table [col sep=comma, x=sigma_sim2, y=OSPA_phi_nocal] {pgf/OSPA_phi.csv};
\addlegendentry{no cal}

\addplot [green, thick, dashed, mark options={style=solid}] 
    table [col sep=comma, x=sigma_sim2, y=OSPA_phi_optcal] {pgf/OSPA_phi.csv};
\addlegendentry{optimal}

\end{axis}
\end{tikzpicture}

%% file: pgf/K_hats.tex
\begin{tikzpicture}
\begin{axis}[
    width=\figurewidth,
    height=\figureheight,
    grid=both,
    xmode=log,
    xmin=9e-3, xmax=1e0,
    xlabel={$\sigma_{\text{w,sim}}$},
    ylabel={$\hat{K}$},
    legend style={font=\small, at={(0.05,0.95)}, anchor=north west},
]

\addplot[
    blue,
    thick,
    mark=square*
]
table[
    col sep=comma,
    x=sigma_sim2,
    y=cal
]{pgf/K_hats.csv};
\addlegendentry{cal}

\addplot[
    red,
    thick,
    dashed,
    mark=o
]
table[
    col sep=comma,
    x=sigma_sim2,
    y=nocal
]{pgf/K_hats.csv};
\addlegendentry{no cal}

\end{axis}
\end{tikzpicture}

%% file: pgf/mean_estimation_times.tex
\begin{tikzpicture}
\begin{axis}[
      width=\figurewidth,
    height=\figureheight,
    grid=both,
    xmode=log,
    xmin=9e-3, xmax=1e0,
    xlabel={$\sigma_{\text{w,sim}}$},
    ylabel={$t_{conv}$ (s)},
    legend style={font=\small, at={(0.05,0.95)}, anchor=north west},
]

\addplot[
    blue,
    thick,
    mark=square*
]
table[
    col sep=comma,
    x=sigma_sim2,
    y=cal
]{pgf/mean_estimation_times.csv};
\addlegendentry{cal}

\addplot[
    red,
    thick,
    dashed,
    mark=o
]
table[
    col sep=comma,
    x=sigma_sim2,
    y=nocal
]{pgf/mean_estimation_times.csv};
\addlegendentry{no cal}

\end{axis}
\end{tikzpicture}